\documentstyle{ioplppt}

\def\bfl{\begin{flushleft}}
\def\efl{\end{flushleft}}
\def\bfr{\begin{flushright}}
\def\efr{\end{flushright}}
\def\bc{\begin{center}}
\def\ec{\end{center}}
\def\be{\begin{equation}}
\def\ee{\end{equation}}
\def\ba{\begin{eqnarray}}
\def\ea{\end{eqnarray}}
\def\nn{\nonumber }
\def\lb#1{\label{#1}}

\def\text#1{\mbox{#1}}
\def\drm{\text{d}}

\def\kevfcub{\;\text{keV} \, \text{fm}^{-3}}
\def\tevfcub{\;\text{TeV} \, \text{fm}^{-3}}
\def\Adequa{\Longleftrightarrow}
\def\Sign#1{\, \text{sign}\left[#1\right] }
\def\Po{Poincar\'e~}
\def\RN{Reissner-Nordstr\"om~}

\begin{document}
\jl{6}
\title{{\small Class. Quant. Grav. 16 (1999) 1737-1744\\
gr-qc/9707054}\\ 
~\\
Radiation fluid singular hypersurfaces with de Sitter interior 
as models of charged extended particles in general relativity}[Radiation 
fluid singular hypersurfaces]

\author{Konstantin G. Zloshchastiev}

\address{Metrostroevskaya 5/453, Dnepropetrovsk 320128, Ukraine}

\begin{abstract}
In present paper we construct the classical and minisuperspace quantum 
models of an extended charged particle.
The modelling is based on the radiation fluid singular hypersurface 
filled with physical vacuum.
We demonstrate that both at classical and quantum levels 
such a model can have equilibrium states
at the radius equal to the classical radius of a charged particle.
In the cosmological context the model could be considered also as
the primary stationary state, having the huge internal energy being
nonobservable for an external observer, from which the Universe 
was born by virtue of the quantum tunnelling.
\end{abstract}

\section{Introduction}

About one hundred years ago, Lorentz suggested a model of an extended 
electron as the body having pure charge and no matter \cite{roh}.
His model had severe problems with stability because the electric repulsion
should eventually lead to explosion of the configuration proposed.
Since that time many modifications of this model have been done.
To provide stability of the Lorentz model, \Po introduced the stresses that 
greatly improved situation \cite{poi}.
After appearance of general relativity Einstein raised the 
particle problem on the relativistic level.
Subsequently, the Lorentz-\Po ideas obtained 
new sounding in terms of repulsive gravitation \cite{gro}, 
and interest to such models has been renewed, especially in connection
with the modelling of exotic extended particles.
In spite of these relativistic models seem to be appropriate 
for a charged
{\it spinless} particle (e.g., $\pi$-meson) rather then for an electron 
(except the works \cite{il}), they nevertheless contain some consistent 
descriptions and concepts;
it is implicitly assumed that the presence of 
spin entails only small deviations from spherical symmetry.

The main idea of the most popular models is to consider the
(external) \RN solution of the Einstein-Maxwell field equations,
\be
\drm s^2 =- 
\left(1-\frac{2 M}{r} + \frac{Q^2}{r^2} \right) \drm t^2 + 
\left(1-\frac{2 M}{r} + \frac{Q^2}{r^2} \right)^{-1} \drm r^2 +
r^2 \drm \Omega^2                                                \lb{eq1}  
\ee
(we will work in terms of gravitational units), and to match it with some
internal one; thereby this matching is performed
across the boundary surface (which is considered as a particle surface), 
i.e., the first and second quadratic forms are continuous on it 
(it is the well-known Lichnerowicz-Darmois junction 
conditions) \cite{gau}.
At the same time, more deep investigations of the electromagnetic 
mass theories, based
on such a matching, elicit a lot of the features which point out
some imperfection of the particle models based on the 
boundary surfaces (i.e., on the discontinuities of first kind).

The first of them is an accumulation of the electric charge on a core
boundary.
Let us illustrate it for the model \cite{cc}.
Cohen and Cohen matched the external \RN solution (\ref{eq1}), at 
$r > R$ where $R$ is a core radius, with the de Sitter one, at $r\leq R$,
\be
\drm s^2 =- 
(1-\lambda^2 r^2)\, \drm t^2 + 
(1-\lambda^2 r^2)^{-1} \drm r^2 +
r^2 \drm \Omega^2, 
\quad 
\lambda^2 = \frac{8 \pi \varepsilon_v}{3},                          \lb{eq2}  
\ee
thereby they obtained 
$M = (4 \pi \varepsilon_v/3) R^3 + Q^2/2 R$, 
$\varepsilon_v$ is the physical vacuum energy density; 
the static chart describes interior at $\lambda r<1$.
The stress-energy tensor components for this solution look like those
for the polarised vacuum \cite{zel}:
\ba
T^0_0 = \varepsilon_v = ~^{(3)}\!\rho + Q^2(r)/8 \pi r^4, \nn\\
T^1_1 = \varepsilon_v = -~^{(3)}\!p + Q^2(r)/8 \pi r^4, \lb{eq3}\\
T^2_2 =T^3_3 = 
\varepsilon_v = - ~^{(3)}\!p_\bot - Q^2(r)/8 \pi r^4, \nn
\ea
where $Q(r)$, $^{(3)}\!\rho$, $^{(3)}\!p$ are the charge, energy density,
and pressure inside $r$, respectively.
The generalisation of the Tolman-Oppenheimer-Volkov equation for 
such charged models yields \cite{trk}
\be
\frac{\drm }{\drm r} ~^{(3)}\!p =  
\frac{1}{8 \pi r^4} \frac{\drm }{\drm r} Q^2(r)
\quad
r\leq R,                                                          \lb{eq4}
\ee
hence
\be
Q(r) = Q\, \theta (r - R),                                          \lb{eq5}
\ee
where $\theta (r)$ is the Heaviside step function.
Thus, there is no electric field inside the core, and charge is 
accumulated on the boundary only.
However, the discontinuities of first kind cannot have surface charge 
distribution {\it a priori}.
The second feature is the energy of the false vacuum is transferred 
to a boundary as well \cite{col,cg}.
Nevertheless,
the discontinuities of first kind cannot have surface energy 
density too.
Such discordances of the boundary-surface models compel to seek 
for the more realistic models, e.g., having non-trivial surface 
properties.

\section{Classical model}

We can consider the system of 
the Einstein-Maxwell field equations plus singular (infinitely thin) 
shell instead of the simple boundary surface \cite{dau,isr,mtw}.
The singular surface has to be the discontinuity of second kind 
(the first quadratic form
is continuous across it but the second one has a finite jump), and,
unlike the boundary surface, can have proper surface charge and
stress-energy tensor
\be
S_{ab}=\sigma u_a u_b + p (u_a u_b +~ ^{(3)}\!g_{ab}),            \lb{eq6}
\ee
where $\sigma$ and $p$ are respectively the surface energy density and 
pressure, $u^a$ is the timelike unit tangent vector, 
$^{(3)}\!g_{ab}$ is the 3-metric on a shell.

Thus, let us study the spherical singular shell filled with de Sitter 
vacuum $\Sigma^-$ (\ref{eq2}) and inducing the external \RN spacetime
$\Sigma^+$ (\ref{eq1}).
The metric of the (2+1)-dimensional spacetime $\Sigma$ 
of the shell can be written in 
terms of the shell's proper time $\tau$ as
\be
^{(3)}\!\drm s^2 = - \drm \tau^2 + R^2 \drm \Omega^2,             \lb{eq7}
\ee
where $R=R(\tau)$ turns to be the proper radius of the shell.

Then the foliated Einstein-Maxwell field equations give the
two equation groups.
The first one is the Einstein-Maxwell equations on the shell, integrability
conditions of which yield both the conservation law of shell's matter
\be
\drm \left( \sigma ~^{(3)}\!g \right) +
p~ \drm \left( ~^{(3)}\!g \right) 
+ ~^{(3)}\!g~  \Delta T^{\tau n}\, \drm \tau =0,                   \lb{eq8}
\ee
where $\Delta T^{\tau n} = (T^{\tau n})^+ - (T^{\tau n})^-$,  
$T^{\tau n}=T^{\alpha\beta} u_\alpha n_\beta$ is the
projection of the stress-energy tensors in the $\Sigma^\pm$
spacetimes on the tangent and normal vectors, 
$^{(3)}\!g=\sqrt{- \det{(^{(3)}\!g_{ab})}} = R^2 \sin{\theta}$ 
(it should be noted that $T^{\tau n} \equiv 0$ 
for the spacetimes (\ref{eq1}), (\ref{eq2})),
and electric charge conservation law which in our case can be reduced to
the relation
$
Q=\text{constant}.
$
The second group includes the equations of motion of a shell which are 
the Lichnerowicz-Darmois-Israel junction conditions 
\be
(K^a_b)^+ - (K^a_b)^- = 4 \pi\sigma (2 u^a u_b + \delta^a_b),
                                                                   \lb{eq9}
\ee
where $(K^a_b)^\pm$ are the extrinsic curvatures for spacetimes 
$\Sigma^\pm$ respectively.
Besides, the associated electromagnetic potential vector 
is discontinuous across the shell.
Following aforesaid, the electromagnetic potential is zero 
inside the shell whereas outside it has to be
\ba
A_i = (-Q/r,0,0,0). \nn
\ea

Thus, taking into account 
(\ref{eq1}), (\ref{eq2}), (\ref{eq7}), and (\ref{eq8}),
the $\theta\theta$ and $\tau\tau$ components of the equation (\ref{eq9}) 
yield
\ba
\epsilon_+ \sqrt{\dot R^2 + 1 - \frac{2 M}{R}+\frac{Q^2}{R^2}}
-
\epsilon_- \sqrt{\dot R^2 + 1 - \lambda^2 R^2} =
-4\pi\sigma R,                                                 \lb{eq10}\\
\frac{R^3 \ddot R + M R -Q^2}
     {
      \epsilon_+ R^3 \sqrt{\dot R^2 + 1 - \frac{2 M}{R}+\frac{Q^2}{R^2}}
     }
-
\frac{\ddot R -\lambda^2 R}
     {
      \epsilon_- \sqrt{\dot R^2 + 1 - \lambda^2 R^2}
     }
=-4\pi \frac{\drm (\sigma R)}{\drm R},                            \lb{eq11}
\ea
where $\epsilon_\pm = \Sign{(K_{\theta\theta})^\pm}$, 
$\dot R = \drm R/\drm \tau$ etc.
The root sign $\epsilon = +1$ if $R$ increases in 
the outward normal of the shell, and $\epsilon = -1$ if $R$ decreases.
Thus, only under the additional condition $\epsilon_+ = \epsilon_-=1$ we 
have the ordinary shell.
Below we will deal with such shells only.

Besides, independently of the Einstein equations 
the equation of state for matter on the shell $p=p(\sigma)$
should be added as well.
The simplest equation of state we choose is the linear one of a barotropic 
fluid \cite{bkkt},
\be
p=\eta \sigma,                                                  \lb{eq12}
\ee
including as a private case, dust ($\eta = 0$), radiation fluid
($\eta = 1/2$, the reduction of shell's spacetime dimensionality is 
taken into account), 
bubble matter ($\eta = - 1$), and so on \cite{bkkt,zlo002,cg}.
Then the constant $\eta$ remains to be arbitrary and will be specified 
below for physical reasons.
It can easily be seen that the equations (\ref{eq8}), (\ref{eq10}), and 
(\ref{eq12}) form a complete system.
Solving equations (\ref{eq8}) and (\ref{eq12}) together, we obtain
\be
\sigma = \frac{\alpha}{4 \pi} R^{-2(\eta+1)},                     \lb{eq13}
\ee
where $\alpha$ is an integration constant related to the surface energy 
density at some fixed $R$, $\alpha>0$ for ordinary shells.
Then the equations (\ref{eq10}) and (\ref{eq11}) can be rewritten more
strictly:
\ba
\sqrt{\dot R^2 + 1 - \frac{2 M}{R}+\frac{Q^2}{R^2}}
- \sqrt{\dot R^2 + 1 - \lambda^2 R^2} =
-\frac{\alpha}{R^{2\eta+1}},                                  \lb{eq14}\\
\frac{R^3 \ddot R + M R -Q^2 }
     {
      R^3 \sqrt{\dot R^2 + 1 - \frac{2 M}{R}+\frac{Q^2}{R^2}}
     }
-
\frac{\ddot R -\lambda^2 R}
     {
      \sqrt{\dot R^2 + 1 - \lambda^2 R^2}
     }
=\frac{\alpha (2\eta +1)}{R^{2(\eta+1)}}.                      \lb{eq15}
\ea
Now we consider this system in the equilibrium $\dot R=\ddot R=0$ at
$R=R_p$ that will correspond to a classical particle radius.
Then these equations turn to be the equilibrium conditions \cite{blp}
\ba
\sqrt{1 - \frac{2 M}{R_p}+\frac{Q^2}{R_p^2}}
- \sqrt{1 - \lambda^2 R_p^2} =
-\frac{\alpha}{R_p^{2\eta+1}},                               \lb{eq16}\\
\frac{M R_p -Q^2}
     {
      R_p^3 \sqrt{1 - \frac{2 M}{R_p}+\frac{Q^2}{R_p^2}}
     }
+
\frac{\lambda^2 R_p}
     {
      \sqrt{1 - \lambda^2 R_p^2}
     }
=\frac{\alpha (2\eta +1)}{R_p^{2(\eta+1)}}.                    \lb{eq17}
\ea
Further, it is well-known that the classical radius $R_c$ of a charged
particle can be defined as the radius at which the function $1-2M/R+Q^2/R^2$ 
approaches a minimum.
Therefore $R_c = Q^2/M$; assuming $R_p = R_c$ we have respectively
\ba
\sqrt{1 - (M/Q)^2} - 
\sqrt{1 - \left( \lambda_c Q^2 / M \right)^2} =
-\alpha_c \left( M/Q^2 \right)^{2\eta+1},                 \lb{eq18}\\
\lambda_c^2
=\alpha_c (2\eta + 1) \left( M /Q^2 \right)^{2\eta+3}
\sqrt{1 - \left( \lambda_c Q^2/M \right)^2}.                    \lb{eq19}
\ea
It should be noted that from the last equation the restriction
for $\eta$, $\eta>-1/2$, follows which appears to be a necessary condition
for existence of equilibrium at $R=R_c$.
Hence it is easy to see that our thin-wall model cannot be ``made'' from
the bubble matter $\eta =-1$ ($\sigma =\text{constant})$.
But the {\it ultrarelativistic} radiation fluid, on the contrary, 
seems to be a very suitable candidate for the role of two-dimensional
matter on the particle surface.
In this connection it is interesting to note that radiation fluid singular
configurations can not be in equilibrium as a rule.
However, in the present case the internal physical vacuum compensates
both the radial pressure caused by the shell and repulsive electrical forces
that stabilises the system. 

Considering the choice $\eta =1/2$ 
and eliminating $\alpha_c$ from the system (\ref{eq18}), (\ref{eq19}), 
we obtain after rejecting a superfluous
root the exact electromagnetic mass relation 
\be
\lambda_c^2=\frac{8\pi\varepsilon_v}{3} =
\frac{2}{9} \left(\frac{M}{Q^2}\right)^2
\left[
2+ \frac{M^2}{Q^2} -
\sqrt{
      \left(1-\frac{M^2}{Q^2} \right)
      \left(4-\frac{M^2}{Q^2} \right)
     }
\right],                                                 \lb{eq20}
\ee
which determines the necessary stabilising energy 
of de Sitter vacuum inside a charged particle within frameworks 
of the classical model having a radiation fluid singular surface.
We can simplify this relation using the fact that the value $M/Q$
turns to be very small for the known elementary particles.
Indeed, performing the Taylor expansion of equation (\ref{eq20}), 
we obtain
\be                                     \lb{eq21}
\varepsilon_v =
\frac{3}{16\pi} \frac{M^4}{Q^6} +
O \left( \frac{M^6}{Q^8} \right).
\ee

Let us perform some numerical estimations.
For instance, for an electron we have: 
$M=0.511 \,\text{MeV} = 6.76 \times 10^{-58}\, \text{m}$,
$Q=1.38\times 10^{-36} \, \text{m}$,
$R_c=2.8\times 10^{-15} \, \text{m}$, hence
$\varepsilon_v=2.5 \kevfcub$.
For  $\pi^\pm$-mesons we get: 
$M=139.57 \,\text{MeV}$,
$Q=1.38\times 10^{-36} \, \text{m}$,
$R_c=1.02\times 10^{-17} \, \text{m}$, hence
$\varepsilon_v=14 \tevfcub$.
One can see that for particles of reasonably large mass 
the huge internal vacuum energy and electrical charge can be 
equilibrated by the surface gravitational forces
providing the (classical) stability of the model.
The gravitational defect of masses causes the property that
the external (observable) mass of a particle appears to be
much less than the internal energy.
The more detailed analysis of the possible applicability of such a
phenomenon will be given below at the consideration of quantum
fluctuations, all the more so latter enforce this feature.

\section{Minisuperspace quantization}

Following the Wheeler-DeWitt's approach in quantum cosmology the 
Universe is considered quantum mechanically and is described by a wave
function \cite{dew}. 
This function is defined on the superspace which is the space of all
admissible metrics and accompanying fields.
The minisuperspace approach appears to be the direct application of
Wheeler-DeWitt's quantization procedure for (2+1)-dimensional singular
hypersurfaces having own internal 
three-metric $^{(3)}\!g_{ab}$ \cite{vil,zlo001}.
Thereby, in spherically symmetric case the world sheet of a singular
hypersurface is determined by a single function, viz.,
proper radius $R(\tau)$.
In the absence of a unified rigorous axiomatic approach \cite{not} 
this method has the following advantages in comparison with others:
(i) it is simple and gives heuristic results in most cases
in a nonperturbative way, which is very important for non-linear
general relativity,
(ii) there explicitly exists conformity with the correspondence 
principle that improves physical interpretation of all the
concepts of the theory. 

Let us consider a minisuperspace model initially
described by the Lagrangian
\be                                     \lb{eq22}
L = \frac{m \dot R^2}{2} - U,
\ee
where
\be                                     \lb{eq23}
U = U_1 + U_2,
\ee
\[
U_1 = 
-\frac{\lambda^4 R^7}{8 \alpha} + \frac{M \lambda^2 R^4}{2 \alpha}
- \frac{Q^2 \lambda^2 R^3}{4 \alpha} - 
\frac{(\lambda^2 \alpha^2 + 2 M^2)R}{4\alpha},
\]
\[
U_2 = 
\frac{M Q^2}{2\alpha} + \frac{4\alpha^2 - Q^4}{8 \alpha R}
- \frac{\alpha M}{2 R^2} + \frac{\alpha Q^2}{4 R^3} - 
\frac{\alpha^3}{8 R^5},
\]
and
\be                                     \lb{eq24}
m = 4\pi\sigma R^2 = \alpha / R
\ee
is the mass of the radiation fluid shell in the static reference frame.
The equation of motion is therefore
\be                                     \lb{eq25}
\frac{\drm}{ \drm \tau} (m \dot R) =
\frac{m_{,R} \dot R^2}{2} - (U_1 + U_2)_{,R},
\ee
where ``$,R$'' means the derivative with respect to shell radius.
Considering time symmetry, we can easily decrease an order of this
differential equation and obtain
\be                                     \lb{eq26}
\dot R^2 =
\frac{2 R}{\alpha} (H - U_1 - U_2),
\ee
where $H$ is the integration constant.
It can be directly checked that if one supposes $H$ to be vanishing,
\be                                     \lb{eq27}
H = 0,
\ee
then we obtain the (double squared) equation of 
motion of our radiation fluid shell (\ref{eq10}) provided 
eqs. (\ref{eq13}), (\ref{eq24}) and $\eta =1/2$.
Thus, our Lagrangian indeed describes dynamics of 
the charged radiation fluid shell filled with physical vacuum
up to topological features which were described by 
the signs $\epsilon_\pm$.
However, we always can take into account the topology $\epsilon_\pm$
both at classical (rejecting superfluous roots) and quantum 
(considering boundary conditions for the corresponding 
Wheeler-DeWitt equation) levels.

Further, introducing the momentum $\Pi=m \dot R$,
from (\ref{eq27}) we obtain that the (super) Hamiltonian,
\be                                     \lb{eq28}
{\cal H} = \Pi \dot R - L \equiv H,
\ee
has to be equal zero on the real trajectories (\ref{eq10}).
Therefore, we have obtained the constraint
\be                                     \lb{eq29}
\frac{\Pi^2}{2 \alpha/R} + U_1 + U_2 = 0,
\ee
which in quantum case ($\hat\Pi= -i \partial/\partial R$, here and 
below we assume Planckian units)
yields the Wheeler-DeWitt equation for
the wave function $\Psi (R)$:
\be                                     \lb{eq30}
\Psi^{\prime\prime} + \Xi \Psi = 0,
\ee
where
\be                                     \lb{eq31}
\Xi =  \Xi_1 + \Xi_2,
\ee
\[
\Xi_1 = 
\frac{\lambda^4 R^6}{4} - M \lambda^2 R^3 + 
\frac{Q^2 \lambda^2 R^2}{2} + \frac{\lambda^2 \alpha^2}{2} + M^2,
\]
\[
\Xi_2 = 
-\frac{M Q^2}{R} + \frac{Q^4 -4\alpha^2}{4 R^2} 
+ \frac{M \alpha^2}{R^3} - \frac{Q^2 \alpha^2}{2 R^4}
+\frac{\alpha^4}{4 R^6}.
\]

Now we will consider small quantum oscillations of radius
in a neighbourhood of the state of {\it dynamical} equilibrium
represented by the minimum point at $R=R_c$.
In our theory the two types of equilibrium exist: kinematic 
\be                                     \lb{eq32}
\dot R = 0 \Adequa \Xi =0, 
\ee
and
dynamical (all acting forces are equilibrated)
\be                                     \lb{eq33}
\ddot R = 0 \Adequa \Xi^\prime =0.
\ee
In quantum theory the simultaneous exact satisfaction of the 
requirements $R=R_c$ and $\dot R=0$
turns to be forbidden because of the uncertainty principle.
The quantum analogue of classical equilibrium is the (quantum) 
dynamical equilibrium when a system passes to the bound state
characterised by a discrete spectrum for corresponding parameters,
first of all, the total mass (energy) $M$.
Therefore, we will suppose that eq. (\ref{eq33}), 
not (\ref{eq32}), is valid
at $R=R_c$.
Then eqs. (\ref{eq31}), (\ref{eq33}) give
\be                                     \lb{eq34}
\alpha^2 = \frac{Q^6}{3 M^6}
\Bigl[
      M^2 (2 Q^2 - M^2) + \zeta \beta
\Bigr],
\ee
where $\zeta=\Sign{\beta}$ and
\[
\beta = 
\sqrt{ 
      4 M^4 Q^2 (Q^2 -M^2)+ (2 M^4 - 3 \lambda^2 Q^6)^2
     }.
\]
Then in the first-order $\hbar$-expansion approximation from
eqs. (\ref{eq30}), (\ref{eq31}), (\ref{eq33}) and (\ref{eq34})
we obtain the wave equation for some quantum harmonic oscillator of
unit mass
\be                                     \lb{eq35}
\frac{1}{2} \frac{\drm^2 \Psi}{\drm x^2} + 
\left[
      \varepsilon - \frac{\omega^2 x^2}{2} 
\right]
\Psi = 0,
\ee
where $x=R-Q^2/M$,
\ba                                     
2\varepsilon =
\frac{1}{18 M^6}
\Bigl[
      9 \lambda^4 Q^{12} - 6 \lambda^2 Q^6 M^2 (3 M^2-Q^2)  \nn\\
      - 4 M^4 (2 Q^4
      - 2 M^2 Q^2 -M^4)+
      (3\lambda^2 Q^6 - 4 M^2 Q^2 +2 M^4)\zeta\beta
\Bigr],                                             \lb{eq36}
\ea
\ba                                                 
\omega^2 = \frac{1}{3 M^4 Q ^4}
\Bigl[
3 \lambda^2 Q^6 (19 M^4 -18 \lambda^2 Q^6)  
-2 M^4 (8 Q^4 - 9 M^2 Q^2 +7)               \nn\\
-M^2 (8 Q^2 - 5 M^2) \zeta\beta
\Bigr].                                               \lb{eq37}
\ea

Further, for bound states the quantum boundary conditions, 
corresponding to the singular Stourm-Liouville problem,
require $\Psi (-\infty) = \Psi (+\infty) = 0$
(that has to be somewhere artificial but admissible  
$\hbar$-expansion condition if we
suppose the wave function to be localised near the stationary point
of the effective potential (\ref{eq26})),
and the normalised solution of eq. (\ref{eq35}), which approximate
the exact solution in the neighbourhood of the stationary point, 
can be expressed by means of the Hermite polynomials $H_n (y)$ 
\be                                                \label{eq38}
\Psi (y) = 
\left(
      2^n \sqrt{\pi} n\verb|!|   
\right)^{-1/2}
\exp{(-y^2/2)}\, H_n (y),                                   
\ee
where $n=0,\,1,\,2,...\,$, $y=\sqrt{\omega} x$.
The spectrum of $\varepsilon$ is known to be
\be                                                \label{eq39}
\varepsilon = \omega (n+1/2),
\ee
which yields, together with eqs. (\ref{eq36}) and (\ref{eq37}),
the restriction on the parameters $\lambda$, $M$, $Q$ of the 
studied configuration, radiation fluid shell filled with physical 
vacuum, in the bound state.

In general case equation (\ref{eq39}) is hard to solve, however in
squared (\ref{eq39}) we can
perform expansion in series with respect to the 
ratio $\delta=M/Q$ which is
expected to be small for non-macroscopical particles.
Then we obtain (${\cal N} =2 n + 1$):
\be                                                \label{eq40}
\zeta = +1:\> -\frac{1}{2} \lambda^2 Q^2 \delta^{-4}
+\frac{5}{9} \delta^{-3} + O(\delta^{-2}) = 
\left(
\frac{3 {\cal N}}{\lambda Q^3}
\right)^2,
\ee
\be                                                \label{eq41}
\zeta = -1:\> 
-\frac{1}{2} \delta^{-2} 
+ \frac{1+3\lambda^2 Q^2}{3\lambda^2 Q^2} \delta^{-1}
+ O(\delta^0) = 
\left(
\frac{3 {\cal N}}{Q^2}
\right)^2.
\ee
Let us consider, e.g., the case $\zeta = -1$.
Then from the last equation the total mass-energy spectrum follows:
\be                                                \label{eq42}
M=
\frac{
      3 \lambda^2 Q^3
     }
     {
      1+3\lambda^2 Q^2 \pm 
      \sqrt{
            (1+3\lambda^2 Q^2)^2 -162 \lambda^4 {\cal N}^2
           }
     },
\ee
in which the deSitter's density $\lambda^2$ 
remains to be a free parameter.
It appears to be the 
one of the most severe problems if we would seek to 
successively answer the questions: what are the particles our
model describes? are they known or exotic particles? are they really
microscopical objects? and so on.
These questions are open yet.

Finally, let us perform analytical estimations of internal vacuum 
energy density.
Equation (\ref{eq41}) yields the expression 
(the case $\zeta = 1$ gives the same estimate) 
\be                                     \lb{eq43}
\varepsilon_v =
\frac{1}{4\pi} \frac{M}{Q^3} +
O \left( \frac{M^2}{Q^4} \right),
\ee
which evidently does not depend on $n$ at this order.
Comparing expressions (\ref{eq21}) and (\ref{eq43}) one can see that 
in quantum case $\varepsilon_v$ has to be much more 
(three orders with respect to $\delta$) than its classical 
counterpart.
Thus, quantum fluctuations appeared to be the strong additive reason
preserving the model from both the gravitational shrinking and 
electrical explosion.
The gravitational defect of mass enforced by them leads to the internal 
vacuum energy confined inside such a shell\footnote{The 
internal energy density is estimated to be of order 
$10^{-21}-10^{-19}$ of the Planckian one for the trial 
masses of the electron and pion.}
is much more than the external one.
From the cosmological point of view it is an expected result:
there exists the shell models of the birth of the Universe from 
a small region to infinite size by virtue of the quantum tunnelling from
a {\it stationary} state (see \cite{vil,gut} and references therein).
The released huge energy is nothing but the deconfined 
internal energy, the 
internal space turns to be the whole spacetime of the created Universe,
and the surface electrical charge is non-observable for the
internal observer.
In this connection our radiation fluid shell indeed
could be considered also as the primary stationary state because of
the four necessary reasons: 
(i) it has a local stationary state, 
(ii) it has the states with infinite radius (see the effective potential
defined by the equation (\ref{eq26})), 
(iii) the  barrier between them and the stationary state has 
a finite height,
(iv) it demonstrates the opportunity of the presence of huge 
confined energy inside itself.

\section{Conclusion}

Thus, in present paper we have constructed the classical 
model of an extended charged particle
based on the equilibrium radiation 
fluid shell filled with physical vacuum.
Thereby we found the configuration in which the 
radiation fluid can be confined.
Then we 
performed the minisuperspace quantization of this model and obtained
the Wheeler-DeWitt equation.
Resolving it in a neighbourhood of the point of dynamical equilibrium,
we established that the configuration can have
bound states near this point, obtained discrete spectra for them, and
performed quantitative estimations for some trial masses.
Thus it was shown that the radiation 
fluid singular surface filled with physical vacuum indeed can model some
particle-like objects both at classical and quantum levels.
Besides, in the cosmological context the model could be 
considered also as the primary stationary state from which 
the Universe can be born via the quantum tunnelling.

\def\JETPr{Zh. Eksp. Teor. Fiz.}
\def\CMPh{Commun. Math. Phys.}
\def\JPh{J. Phys.}
\def\CJP{Czech. J. Phys.}
\def\FP{Fortschr. Phys.}
\def\LMPh {Lett. Math. Phys.}
\def\MPL {Mod. Phys. Lett.}
\def\NPh  {Nucl. Phys.}
\def\PhE  {Phys.Essays}
\def\PhL  {Phys. Lett.}
\def\PhR  {Phys. Rev.}
\def\PhRL {Phys. Rev. Lett.}
\def\PhRp {Phys. Rep.}
\def\NCim {Nuovo Cimento}
\def\NuPB {Nucl. Phys.}
\def\GRG {Gen. Relativ. Gravit.}
\def\CQG {Class. Quantum Grav.}
\def\prp {report}
\def\Prp {Report}

\def\jn#1#2#3#4#5{#5 {\it #1}{#2} {\bf #3} {#4}}
\def\boo#1#2#3#4#5{#4 {\it #1} ({#3}: {#2}){#5}}

\end{document}